# Three-dimensional topological disclination in acoustic crystals

Zhenxiao Zhu[1†], Yan Meng[2†], Minmiao Wang [3†], Xiang Xi[2], Yuxin Zhong[1], Linyun Yang[4], Bei Yan[5], Jingming Chen[1], Ziyao Wang[1], Thomas Christensen[6], Caigui Jiang[7*], Changqing Xu[8*], Ce Shang[9*], Zhen Gao[1*]

[1]*State Key Laboratory of Optical Fiber and Cable Manufacture Technology, Department of Electronic and Electrical Engineering, Southern University of Science and Technology, Shenzhen 518055, China.*
[2]*School of Electrical Engineering and Intelligentization, Dongguan University of Technology, Dongguan, 523808, China.*
[3]*King Abdullah University of Science and Technology (KAUST), Physical Science and Engineering Division (PSE), Thuwal 23955-6900, Saudi Arabia.*
[4]*College of Aerospace Engineering, Chongqing University, Chongqing, 400030, China.*
[5]*Hubei Province Key Laboratory of Systems Science in Metallurgical Process, and College of Science, Wuhan University of Science and Technology, Wuhan 430081, China.*
[6]*Department of Photonics and Electrical Engineering, Technical University of Denmark, Lyngby 2800, Denmark*
[7]*Institute of Artificial Intelligence and Robotics, Xi'an Jiaotong University, Xi'an 710049, China.*
[8]*Key Laboratory of State Manipulation and Advanced Materials in Provincial Universities, School of Physics and Technology, Nanjing Normal University, Nanjing 210023, China.*
[9]*Aerospace Information Research Institute, Chinese Academy of Sciences, Beijing 100094, China.*
†*These authors contributed equally: Zhenxiao Zhu, Yan Meng, Minmiao Wang*

Topological disclinations, crystallographic defects that break rotation lattice symmetry, have attracted great interest and exhibited wide applications in cavities, waveguides, and lasers. However, topological disclinations have thus far been predominantly restricted to two-dimensional (2D) systems owing to the substantial challenges in constructing such defects in three-dimensional (3D) systems and characterizing their topological features. Here we report the theoretical proposal and experimental demonstration of a 3D topological disclination that exhibits fractional (1/2) charge and zero-dimensional (0D) topological bound states, realized by cutting-and-gluing a 3D acoustic topological crystalline insulator. Using acoustic pump-probe measurements, we directly observe 0D topological disclination states at the disclination core, consistent with the tight-binding model and full-wave simulation results. Our results extend the research frontier of topological disclinations and open a new paradigm for exploring the interplay between momentum-space band topology and the real-space defect topology in 3D and higher dimensions.

In general, topological phenomena in crystalline or artificial materials can be categorized into two distinct categories: real-space topological defects (e.g. Dirac vortex, disclination, dislocation) [1–3] that are robust under local deformations and associated with localized states trapped in a crystalline lattice, and momentum-space band topology [4,5] that classifies different topological phases and gives rise to topologically protected boundary states. Although these two phenomena are typically studied separately, recent advances have revealed that their interplay can give rise to many fascinating phenomena and promising applications [6–8], such as zero-dimensional (0D) robust topological bound states [9–23], one-dimensional (1D) topologically protected defect modes [24–36], fractional charges [37–40], topological pumpings [41], topological Wannier cycles [42], and topological lasers [43–49]. In particular, topological disclinations that disrupt the lattice structure of topological crystalline insulators (TCIs) can not only probe the crystalline bulk topology, extending the celebrated bulk-boundary correspondence [4,5] to a bulk-disclination correspondence [39,40], but also offer wide applications in free-form waveguides [50,51], robust cavities [15–19], vortex nanolasers [48,49], topological vortex transport with orbital angular momentum [28,52,53], and topological photonic crystal fibres [54].

To date, most previous studies of topological disclination states (TDSs) have been restricted to two-dimensional (2D) systems that support 0D topological bound states with topological protection in only two (in-plane) dimensions. Only two experimental demonstrations of TDSs in three-dimensional (3D) systems have been reported: 1D topological vortex states in an acoustic Weyl semimetal [28] and 0D topological pentagon states in a giant $C_{540}$ acoustic metamaterial [20]. The former is based on periodically stacking 2D disclinations, and the latter is formed on the 2D surfaces of a $C_{540}$ metamaterial: thus neither are genuinely 3D topological disclinations and lack topological protection in all three dimensions. More recently, a novel monopole topological mode was experimentally demonstrated in a 3D Dirac semimetal with hedgehog [14], extending Dirac vortex from 2D to 3D systems. A natural question arises: does 3D topological disclination exist that host fractional charges and 0D TDSs?

In this Letter, by cutting and gluing half of a cubic 3D acoustic TCI to itself, we realize a 3D topological disclination featuring a fractional (1/2) charge and 0D topological bound state. By measuring the acoustic pressure distributions, we directly observe 0D TDS tightly localized at the disclination core, agree well with theoretical predictions and simulated results. Our work extends the celebrated bulk-disclination correspondence from 2D to 3D and provides a versatile platform to explore novel physical phenomena and practical applications enabled by topological defects in 3D topological materials.

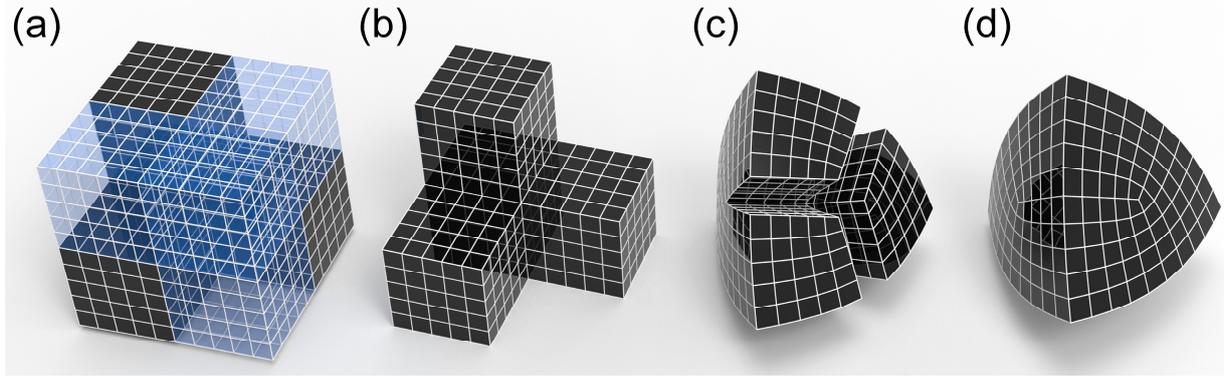

FIG. 1. Constructing a 3D topological disclination by cutting and gluing a cubic lattice. (a) A cubic lattice is divided into eight identical parts and half of them (blue transparent region) are removed. (b) Half of a cubic lattice. All three projection (*xy*, *xz*, and *yz*) planes can form a 2D topological disclination with a Frank angle of $\Omega = -(\pi/2)$. (c) Half of a cubic lattice is gradually contracted and finally glued together, forming a 3D topological disclination at the center with four topological disclination lines emanating from the 3D topological disclination core and outward toward four lattice surface centers [(d)].

*TDSs in a cubic lattice.* We start by constructing a 3D topological disclination in a cubic lattice, as shown in Fig. 1. A perfect cubic lattice is cut into eight identical parts and half of them [blue transparent region in Fig. 1(a)] are removed. Then the remaining parts [Fig. 1(b)] are contracted [Figs. 1(c)] and finally glued together to construct a 3D topological disclination [Fig. 1(d)] at the lattice center. This 3D topological disclination is a type-II disclination [39] since each constituent part comprises complete unit cells. In addition to the 3D topological disclination, this procedure generates four topological disclination lines, emanating from the central 3D topological disclination and outward toward four lattice surface centers, similar to periodically stacked 2D topological disclinations [28].

We present the tight-binding model of a 3D TCI whose intercell couplings are stronger than the intracell couplings ($t_1 > t_2$), as shown in the upper panel of Fig. 2(a) (see Supplementary Note 1 for details). The Wannier centers (red sphere) are located at the corners of the 3D unit cell, as shown in the lower panel of Fig. 2(a). In this configuration, each Wannier center is shared by eight 3D unit cells, contributing a charge of 1/8 to each of them. Figure 2(b) shows the calculated bulk band structure (gray lines) of the 3D TCI which supports three nontrivial band gaps (red region) as well as topological surface, hinge, and corner states in a finite cubic lattice (see Supplemental Material for details). When a cut-and-glue procedure (Fig. 1) deforms the 3D TCI into a 3D topological disclination, the configuration of its Wannier centers is also deformed, as illustrated in Fig. 2c. The 3D topological disclination center (blue tetrahedron) connects four 3D unit cells and each of the four topological disclination lines (green cylinders) connect three 3D unit cells. Jointly, this separates charges associated with the original Wannier centers, resulting in 0D and 1D TDSs in 3D space. The remaining Wannier centers (red spheres) contribute a fractional charge of 1/2 to the 3D topological disclination at the center and a fractional charge of 3/8 to the four topological disclination lines.

Specifically, the fractional charge $Q_{\text{dis}}$ trapped at the 3D topological disclination core can be obtained from the corner charge $Q_{\text{corner}}$ of a symmetrical finite cut-out,

$$Q_{\text{dis}} = mQ_{\text{corner}} \bmod 1 \quad (1)$$

where $m$ is the number of identical constituent parts constructing the 3D topological disclination, $Q_{\text{corner}}$ is the corner charge of a finite symmetrical cut-out which can be calculated from band symmetry data [55–57] (see Supplemental Material for details). Here $m = 4$ and the $Q_{\text{corner}} = 1/8$ for the 3D TCI, such that $Q_{dis} = 1/2$ for the 3D topological disclination (see Supplemental Material for details). Each 3D unit cell, which has a total charge of 1/2 contributed by four Wannier centers, contributes a fractional charge of 1/8 to the 3D topological disclination core which exhibits a fractional charge of 1/2 contributed by four adjacent 3D unit cells, and the remaining fractional charge of 3/8 contributes to three adjacent topological disclination lines. Thus, each topological disclination line exhibits a fractional charge of 3/8 contributed by three adjacent 3D unit cells. This fractional charge configuration [Fig. 2(c)] results in 0D (red dots) and 1D (orange dots) topological disclination states in 3D space, in addition to the 2D surface (green dots), 1D hinge (blue dots), and 0D corner (black dots) states exist in the perfect 3D TCI, as shown in the calculated eigenenergy spectrum in Fig. 2. The spatial distributions of eigenstates of a finite 3D topological disclination can be found in Supplemental Material. For comparison, we construct a trivial insulator by interchanging the intercell and intracell couplings ($t_1 < t_2$), where the Wannier centers (red sphere) are located at the center of the 3D unit cell, as shown in Fig. 2(e). Although this trivial insulator exhibits the same bulk band structure [Fig. 2(f)] with that of the TCI [Fig. 2(b)], its little group representations (see Supplemental Material for details) and charge configuration [Fig. 2(g)] indicates that the charge at the 3D disclination core and disclination lines is 0 since the bulk corner charge vanishes with $Q_{\text{corner}} = 0$ (see Supplemental Material for details). Consequently, no topological disclination or other boundary states exists in the 3D trivial disclination [Fig. 2(h)].

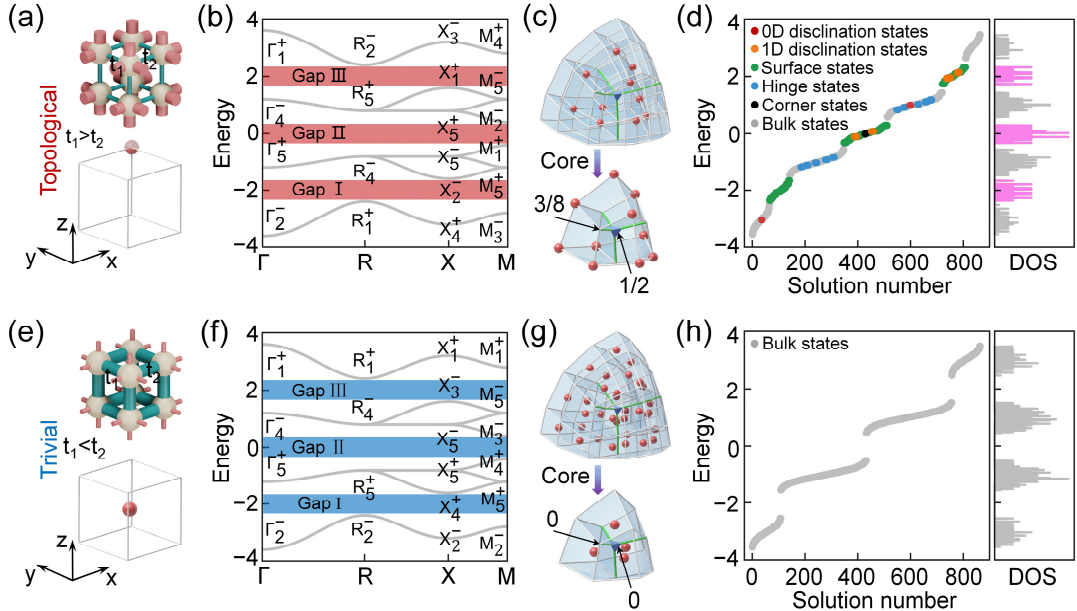

FIG. 2. 3D topological disclination in a cubic lattice. (a) The tight-binding model of a 3D TCI with a cubic lattice and its Wannier center (red sphere) configuration. The green and pink cylinders represent the intracell and intercell couplings, respectively. (b) Bulk band structure of the cubic lattice model in (a). Little group representations are labeled for all bands, which are specified in Supplemental Material. (c) The Wannier center configuration of a finite 3D topological disclination. The lower panel is a zoomed-in view of the disclination core. Each of the four 3D unit cells contributes a fractional charge of 1/8 to the 3D topological disclination, resulting in a fractional charge of 1/2 (mod 1) for 3D topological disclination. In contrast, each of the three 3D unit cells contributes a fractional charge of 1/8 to the topological disclination lines, resulting in a fractional charge of 3/8 (mod 1) for the topological disclination lines. (d) Calculated eigenenergy spectrum and density of state (DOS) of a finite 3D topological disclination. Different topological states are highlighted by different colors and their eigenstate distributions are shown in Supplemental Material. The magenta regions represent the DOS of the in-gap states. (e)-(h) Similar to (a)-(d), but for a tight-binding model of a 3D trivial insulator with a cubic lattice and a 3D disclination. There exist no topological disclination and boundary states. The intercell and intracell couplings of the topological (trivial) tight-binding model are $t_1 = -1$ ($-0.2$) and $t_2 = -0.2$ ($-1$), respectively.

*TDSs in a 3D acoustic crystal.* We now implement the 3D topological disclination in a 3D acoustic crystal [Fig. 3(a)] by mapping the tight-binding model [Fig. 2(c)] with two modifications. First, by adjusting the on-site potentials at the four sites of the 3D topological disclination core [Fig. 3(c)], we can shift three degenerate 0D TDSs (red dots) to the middle band gap from the bulk band. The calculated eigenenergy spectrum of the TDSs as a function of the on-site potential can be found in Supplemental Material. Second, the outermost half of the unit cell of the finite 3D topological disclination was removed to eliminate the topological surface, hinge, and corner states, facilitating experimental observation of the 0D and 1D TDSs (see Supplemental Material for details). Moreover, the 0D and 1D TDSs can also exist in a larger finite 3D topological disclination (see Supplemental Material for details). We then use hollow spherical cavities (beige color) and cylindrical tubes (red and green color) to construct a 3D topological disclination in an acoustic crystal, as shown in Fig. 3(a). A unit cell of a perfect 3D acoustic topological crystalline insulator is shown in Fig. 3(b), where cylindrical tubes with radii of $r_1$ and $r_2$ connecting eight acoustic spherical cavities with radius $r_0$ serve as the intercell and intracell couplings, respectively. Four acoustic cavities with radius $r_3$ at the disclination core [Fig. 3(c)] are adopted to shift the eigenfrequency of 0D TDSs to the middle band gap. The simulated eigenstate spectrum near the middle band gap is shown in Fig. 3(d), in which we can observe 0D TDSs (red dots) in the band gap and 1D TDSs (orange dots) in the bulk. Figures 3(e)-3(g) present the simulated acoustic field distributions of three near-degenerate 0D TDSs, respectively, which are tightly localized at the disclination core. Note that the top quarter of the acoustic crystal is removed to facilitate the visibility of the acoustic pressure distributions at the disclination core. The simulated results of the 1D TDSs can be found in Supplemental Material.

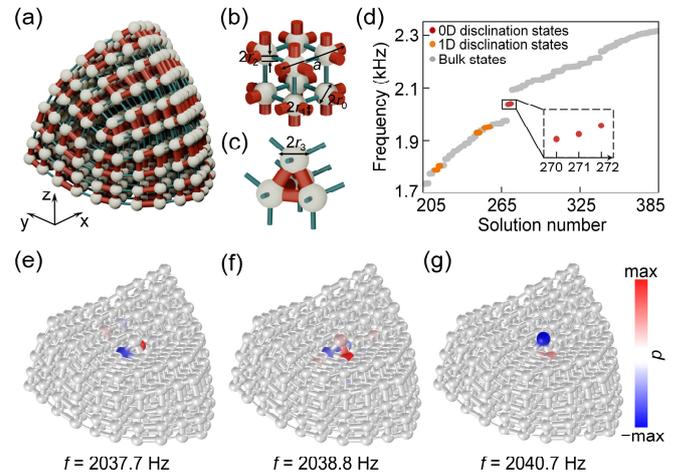

FIG. 3. 3D topological disclination in an acoustic crystal. (a) Schematic of the acoustic crystal with a 3D topological disclination. (b) A unit cell of the cubic acoustic crystal used to construct the 3D topological disclination structure in (a). (c) The zoomed-in view of the disclination core in (a), the radius of the four spherical cavities is enlarged to shift the resonance frequency of the 0D TDSs to the middle band gap. In (a)-(c), The green and red cylinders represent intracell and intercell coupling tubes, respectively. The geometrical parameters are $a = 100$ mm $r_0 = 16$ mm, $r_1 = 8$ mm, $r_2 = 3$ mm, $r_3 = 17.5$ mm, respectively. (d) Simulated eigenstate spectrum of the 3D topological disclination in an acoustic crystal. The red and orange dots represent the 0D and 1D TDSs, respectively. (e)-(f) Simulated acoustic pressure distributions of the 0D TDSs. The simulated acoustic pressure distributions of the 1D TDSs are shown in Supplemental Material.

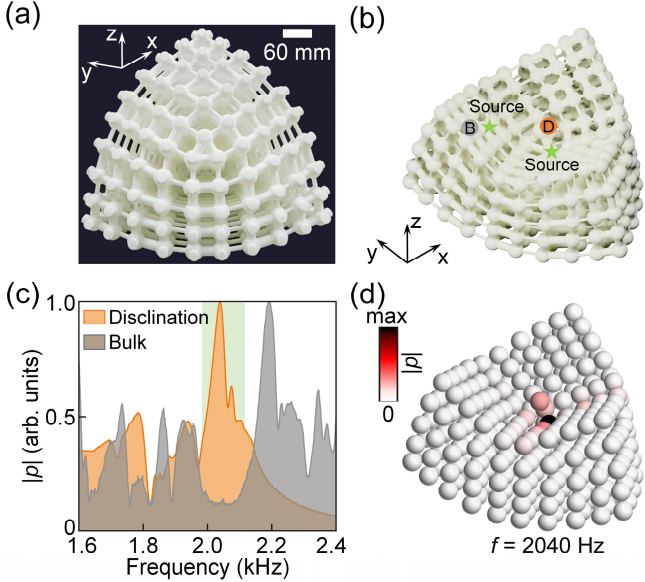

FIG. 4. Experimental demonstration of the 3D topological disclination in an acoustic crystal. (a) Photograph of the fabricated acoustic crystal with a 3D topological disclination. (b) Schematic of the experimental measurements. The green star (orange and gray dots) denotes the acoustic source (probe). (c) Measured spectra of the acoustic pressure amplitude $|p|$ of the disclination (orange color) and bulk (grey color) states with a probe positioned at the disclination center [label D in (b)] and in the bulk region [label B in (b)]. (d) Measured acoustic pressure distribution of the 0D TDS at 2040 Hz.

*Experimental demonstration of 0D TDSs in a 3D acoustic crystal.* Finally, we experimentally demonstrate the 0D TDSs in a 3D acoustic crystal. The experimental sample is shown in Fig. 4(a), which is fabricated by 3D printing technology. To measure the acoustic pressure distributions inside the sample, air holes with a radius of 8 mm are perforated in all cavities on three side surfaces, which are blocked by plugs when not in use. A point-like acoustic source is placed into a cavity near the disclination core (green star) and an acoustic probe is inserted into all cavities to measure the transmission spectra and acoustic pressure distributions (see Methods for more details), as illustrated in Fig. 4(b). Figure 4(c) presents the measured acoustic pressure $|p|$ versus frequency for probes placed at the disclination core (orange color) and in the bulk region (grey color). A bulk bandgap near 2040 Hz (green region) can be observed, within which there exists a resonance peak representing the 0D TDSs, matching well with the simulated results in Fig. 3(d). Figure 4(d) shows the measured acoustic pressure distributions of the 0D TDSs at 2040 Hz, which is tightly confined at the disclination core and consistent with the simulated results in Figs. 3(e)-3(f). The top quarter of the 3D topological disclination is removed for visibility. The measured acoustic pressure distributions of the 1D TDSs can be found in Supplemental Material.

*Conclusion*. We have theoretically proposed and experimentally demonstrated a 3D topological disclination featuring fractional charge and 0D TDSs in a 3D acoustic crystal, extending the bulk-disclination correspondence from 2D to 3D. We show that the fractional charge (1/2) at the 3D topological disclination core can be calculated using reciprocal-space symmetry analysis and understood by the configuration of the Wannier centers. Our work unveils the rich interplay between the real-space topology of topological defects and the reciprocal-space band topology in 3D systems, and may have potential applications in designing 3D topological cavities, lasers, and fibers. Moreover, it will be interesting to explore TDSs with charge fractionalization in 3D non-Hermitian systems [58,59]. We envision that the underlying topological phenomena of 3D topological disclinations can also be extended to other classical wave and condensed matter systems.

Z.G. acknowledges the funding from the National Natural Science Foundation of China under grants No. 62361166627 and 62375118, Guangdong Basic and Applied Basic Research Foundation under grant No.2024A1515012770, Shenzhen Science and Technology Innovation Commission under grants No. 202208151111105001 and 202308073000209, High level of special funds under grant No. G03034K004. Z.Z acknowledges the support from the National Natural Science Foundation of China under grant No. 12404053, Guangdong Basic and Applied Basic Research Foundation under grant No. 2025A1515012284. Y.M acknowledges the support from the National Natural Science Foundation of China under Grant No. 12304484, Guangdong Basic and Applied Basic Research Foundation under grant No. 2024A1515011371. C.X. acknowledges the support from the Jiangsu Specially Appointed Professor Program No. 164080H00244, the Special funds for postdoctoral overseas recruitment, Ministry of Education of China No. 164080H0262405 and Natural Science Foundation of Jiangsu Province No. BK20240576. T.C. acknowledges the support of a research grant (project no. 42106) from Villum Fonden. C.J. acknowledges the support from the National Natural Science Foundation of China under grants No. 62088102. X. X. acknowledges the support from the National Natural Science Foundation of China under grant No.62405053, Guangdong Basic and Applied Basic Research Foundation under grant No. 2025A1515012229.

*Corresponding author.
gaoz@sustech.edu.cn (Z.G.);
shangce@aircas.ac.cn (C.S.);
changqing.xu@nnu.edu.cn (C.X.);
cgjiang@xjtu.edu.cn (C.J.).

[1] N. D. Mermin, *The topological theory of defects in ordered media*, Rev. Mod. Phys. **51**, 591 (1979).

[2] M. Kleman and J. Friedel, *Disclinations, dislocations, and continuous defects: a reappraisal*, Rev. Mod. Phys. **80**, 61 (2008).

[3] J. M. Kosterlitz, *Nobel Lecture: Topological defects and phase transitions*, Rev. Mod. Phys. **89**, 040501 (2017).

[4] M. Z. Hasan and C. L. Kane, *Colloquium: topological insulators*, Rev. Mod. Phys. **82**, 3045 (2010).

[5] X.-L. Qi and S.-C. Zhang, *Topological insulators and superconductors*, Rev. Mod. Phys. **83**, 1057 (2011).

[6] J. C. Y. Teo and C. L. Kane, *Topological defects and gapless modes in insulators and superconductors*, Phys. Rev. B **82**, 115120 (2010).

[7] J. C. Y. Teo and T. L. Hughes, *Topological defects in symmetry-protected topological phases*, Annu. Rev. Condens. Matter Phys. **8**, 211 (2017).

[8] Z.-K. Lin, Q. Wang, Y. Liu, H. Xue, B. Zhang, Y. Chong, and J.-H. Jiang, *Topological phenomena at defects in acoustic, photonic and solid-state lattices*, Nat. Rev. Phys. **5**, 483 (2023).

[9] C. Chen, N. Lera, R. Chaunsali, D. Torrent, J. V. Alvarez, J. Yang, P. San-Jose, and J. Christensen, *Mechanical analogue of a Majorana bound state*, Adv. Mater. **31**, 1904386 (2019).

[10] P. Gao, D. Torrent, F. Cervera, P. San-Jose, J. Sánchez-Dehesa, and J. Christensen, *Majorana-like zero modes in Kekulé distorted sonic lattices*, Phys. Rev. Lett. **123**, 196601 (2019).

[11] X. Gao, L. Yang, H. Lin, L. Zhang, J. Li, F. Bo, Z. Wang, and L. Lu, *Dirac-vortex topological cavities*, Nat. Nanotechnol. **15**, 1012 (2020).

[12] A. J. Menssen, J. Guan, D. Felce, M. J. Booth, and I. A. Walmsley, *Photonic topological mode bound to a vortex*, Phys. Rev. Lett. **125**, 117401 (2020).

[13] J. Noh, T. Schuster, T. Iadecola, S. Huang, M. Wang, K. P. Chen, C. Chamon, and M. C. Rechtsman, *Braiding photonic topological zero modes*, Nat. Phys. **16**, 989 (2020).

[14] H. Cheng, J. Yang, Z. Wang, and L. Lu, *Observation of monopole topological mode*, Nat. Commun. **15**, 7327 (2024).

[15] A. Rüegg and C. Lin, *Bound states of conical singularities in graphene-based topological insulators*, Phys. Rev. Lett. **110**, 046401 (2013).

[16] J. C. Y. Teo and T. L. Hughes, *Existence of Majorana-Fermion bound states on disclinations and the classification of topological crystalline superconductors in two dimensions*, Phys. Rev. Lett. **111**, 047006 (2013).

[17] Y. Deng, W. A. Benalcazar, Z.-G. Chen, M. Oudich, G. Ma, and Y. Jing, *Observation of degenerate zero-energy topological states at disclinations in an acoustic lattice*, Phys. Rev. Lett. **128**, 174301 (2022).

[18] Y. Chen, Y. Yin, Z.-K. Lin, Z.-H. Zheng, Y. Liu, J. Li, J.-H. Jiang, and H. Chen, *Observation of topological p-orbital disclination states in non-Euclidean acoustic metamaterials*, Phys. Rev. Lett. **129**, 154301 (2022).

[19] B. Xia, Z. Jiang, L. Tong, S. Zheng, and X. Man, *Topological bound states in elastic phononic plates induced by disclinations*, Acta Mech. Sin. **38**, 521459 (2022).

[20] D. Liao, J. Zhang, S. Wang, Z. Zhang, A. Cortijo, M. A. H. Vozmediano, F. Guinea, Y. Cheng, X. Liu, and J. Christensen, *Visualizing the topological pentagon states of a giant $C_{540}$ metamaterial*, Nat. Commun. **15**, 9644 (2024).

[21] F.-F. Li, H.-X. Wang, Z. Xiong, Q. Lou, P. Chen, R.-X. Wu, Y. Poo, J.-H. Jiang, and S. John, *Topological light-trapping on a dislocation*, Nat. Commun. **9**, 2462 (2018).

[22] F. Schindler, S. S. Tsirkin, T. Neupert, B. Andrei Bernevig, and B. J. Wieder, *Topological zero-dimensional defect and flux states in three-dimensional insulators*, Nat. Commun. **13**, 5791 (2022).

[23] Y. Liu, X. Zhang, C.-P. Liang, F.-F. Li, Y. Poo, and J.-H. Jiang, *Observation of bulk-dislocation correspondence in photonic crystals*, Phys. Rev. Appl. **22**, 064097 (2024).

[24] H. Lin and L. Lu, *Dirac-vortex topological photonic crystal fibre*, Light Sci. Appl. **9**, 202 (2020).

[25] L. Lu, H. Gao, and Z. Wang, *Topological one-way fiber of second Chern number*, Nat. Commun. **9**, 5384 (2018).

[26] J. Ma, D. Jia, L. Zhang, Y. Guan, Y. Ge, H. Sun, S. Yuan, H. Chen, Y. Yang, and X. Zhang, *Observation of vortex-string chiral modes in metamaterials*, Nat. Commun. **15**, 2332 (2024).

[27] H.-S. Lai, X.-H. Gou, C. He, and Y.-F. Chen, *Topological phononic fiber of second spin-Chern number*, Phys. Rev. Lett. **133**, 226602 (2024).

[28] Q. Wang, Y. Ge, H. Sun, H. Xue, D. Jia, Y. Guan, S. Yuan, B. Zhang, and Y. D. Chong, *Vortex states in an acoustic Weyl crystal with a topological lattice defect*, Nat. Commun. **12**, 3654 (2021).

[29] Y. Ran, Y. Zhang, and A. Vishwanath, *One-dimensional topologically protected modes in topological insulators with lattice dislocations*, Nat. Phys. **5**, 298 (2009).

[30] H. Sumiyoshi and S. Fujimoto, *Torsional chiral magnetic effect in a Weyl semimetal with a topological defect*, Phys. Rev. Lett. **116**, 166601 (2016).

[31] H. Xue, D. Jia, Y. Ge, Y. Guan, Q. Wang, S. Yuan, H. Sun, Y. D. Chong, and B. Zhang, *Observation of dislocation-induced topological modes in a three-dimensional acoustic topological insulator*, Phys. Rev. Lett. **127**, 214301 (2021).

[32] L. Ye, C. Qiu, M. Xiao, T. Li, J. Du, M. Ke, and Z. Liu, *Topological dislocation modes in three-dimensional acoustic topological insulators*, Nat. Commun. **13**, 508 (2022).

[33] E. Lustig, L. J. Maczewsky, J. Beck, T. Biesenthal, M. Heinrich, Z. Yang, Y. Plotnik, A. Szameit, and M. Segev, *Photonic topological insulator induced by a dislocation in three dimensions*, Nature **609**, 931 (2022).

[34] X.-D. Chen, F.-L. Shi, J.-W. Liu, K. Shen, X.-T. He, C. T. Chan, W.-J. Chen, and J.-W. Dong, *Second Chern crystals with inherently non-trivial topology*, Natl. Sci. Rev. **10**, nwac289 (2023).

[35] Y. Wu, Z.-K. Lin, Y. Yang, Z. Song, F. Li, and J.-H. Jiang, *Probing fragile topology with dislocations*, Sci. Bull. **69**, 3657 (2024).

[36] Y. Zhou, R. Davis, L. Chen, E. Wen, P. Bandaru, and D. Sievenpiper, *Helical Phononic Modes Induced by a Screw Dislocation*, arXiv:2404.18347.

[37] E. Lee, A. Furusaki, and B.-J. Yang, *Fractional charge bound to a vortex in two-dimensional topological crystalline insulators*, Phys. Rev. B **101**, 241109 (2020).

[38] T. Li, P. Zhu, W. A. Benalcazar, and T. L. Hughes, *Fractional disclination charge in two-dimensional $C_n$-symmetric topological crystalline insulators*, Phys. Rev. B **101**, 115115 (2020).

[39] C. W. Peterson, T. Li, W. Jiang, T. L. Hughes, and G. Bahl, *Trapped fractional charges at bulk defects in topological insulators*, Nature **589**, 376 (2021).

[40] Y. Liu, S. Leung, F.-F. Li, Z.-K. Lin, X. Tao, Y. Poo, and J.-H. Jiang, *Bulk–disclination correspondence in topological crystalline insulators*, Nature **589**, 381 (2021).

[41] B.-Y. Xie, O. You, and S. Zhang, *Photonic topological pump between chiral disclination states*, Phys. Rev. A **106**, L021502 (2022).

[42] Z.-K. Lin, Y. Wu, B. Jiang, Y. Liu, S.-Q. Wu, F. Li, and J.-H. Jiang, *Topological Wannier cycles induced by sub-unit-cell artificial gauge flux in a sonic crystal*, Nat. Mater. **21**, 430 (2022).

[43] L. Yang, G. Li, X. Gao, and L. Lu, *Topological-cavity surface-emitting laser*, Nat. Photonics **16**, 279 (2022).

[44] S. Han et al., *Photonic Majorana quantum cascade laser with polarization-winding emission*, Nat. Commun. **14**, 707 (2023).

[45] J. Ma et al., *Room-temperature continuous-wave topological Dirac-vortex microcavity lasers on silicon*, Light Sci. Appl. **12**, 255 (2023).

[46] J. Liu et al., *High-power electrically pumped terahertz topological laser based on a surface metallic Dirac-vortex cavity*, Nat. Commun. **15**, 4431 (2024).

[47] X. Xi, J. Ma, and X. Sun, *A topological parametric phonon oscillator*, Adv. Mater. **37**, 2309015 (2025).

[48] M.-S. Hwang, H.-R. Kim, J. Kim, B.-J. Yang, Y. Kivshar, and H.-G. Park, *Vortex nanolaser based on a photonic disclination cavity*, Nat. Photonics **18**, 286 (2024).

[49] F. Jin, S. Mandal, X. Wang, B. Zhang, and R. Su, *Perovskite Topological Exciton-Polariton Disclination Laser at Room Temperature*, arXiv:2404.18360.

[50] Q. Wang, H. Xue, B. Zhang, and Y. D. Chong, *Observation of protected photonic edge states induced by real-space topological lattice defects*, Phys. Rev. Lett. **124**, 243602 (2020).

[51] Y. Wang, Y. Ge, Y.-J. Lu, S. Gu, Y.-J. Guan, Q.-R. Si, S.-Q. Yuan, Q. Wang, H.-X. Sun, and H. Xue, *Free-form acoustic topological waveguides enabled by topological lattice defects*, Phys. Rev. B **109**, L180101 (2024).

[52] C. Huang, C. Shang, Y. V. Kartashov, and F. Ye, *Vortex solitons in topological disclination lattices*, Nanophotonics **13**, 3495 (2024).

[53] Z. Hu, D. Bongiovanni, Z. Wang, X. Wang, D. Song, J. Xu, R. Morandotti, H. Buljan, and Z. Chen, *Topological orbital angular momentum extraction and twofold protection of vortex transport*, Nat. Photonics (2024).

[54] B. Zhu et al., *Topological Photonic Crystal Fibre*, arXiv:2501.15107.

[55] B. Bradlyn, L. Elcoro, J. Cano, M. G. Vergniory, Z. Wang, C. Felser, M. I. Aroyo, and B. A. Bernevig, *Topological quantum chemistry*, Nature **547**, 298 (2017).

[56] Y. Fang and J. Cano, *Filling anomaly for general two- and three-dimensional $C_4$ symmetric lattices*, Phys. Rev. B **103**, 165109 (2021).

[57] K. Naito, R. Takahashi, H. Watanabe, and S. Murakami, *Fractional hinge and corner charges in various crystal shapes with cubic symmetry*, Phys. Rev. B **105**, 045126 (2022).

[58] R. Banerjee, S. Mandal, Y. Y. Terh, S. Lin, G.-G. Liu, B. Zhang, and Y. D. Chong, *Topological disclination states and charge fractionalization in a non-Hermitian lattice*, Phys. Rev. Lett. **133**, 233804 (2024).

[59] R. Li et al., *Observation of Non-Hermitian Topological Disclination States and Charge Fractionalization*, arXiv:2502.04922.